\documentclass{nature}
\usepackage[english]{babel}
\usepackage{amsmath}
\usepackage{amsfonts}
\usepackage{amssymb}
\usepackage{graphicx}
\usepackage{pslatex}
\usepackage{float}

\usepackage[labelformat=empty]{caption}

\let\oldthebibliography=\thebibliography
\let\oldendthebibliography=\endthebibliography
\renewenvironment{thebibliography}[1]{%
    \oldthebibliography{#1}%
    \setcounter{enumiv}{29 }%
}{\oldendthebibliography}
    
\title{A large light-mass component of cosmic rays at 10$^{17}$ - 10$^{17.5}$ eV from radio observations}
\begin{document}

\maketitle
\author{
S.~Buitink$^{1,2}$, 
A.~Corstanje$^{2}$, 
H.~Falcke$^{2,3,4,5}$, 
J.~R.~H\"orandel$^{2,4}$, 
T.~Huege$^{6}$, 
A.~Nelles$^{2,7}$, 
J.~P.~Rachen$^{2}$, 
L.~Rossetto$^{2}$, 
P~.Schellart$^{2}$, 
O.~Scholten$^{8,9}$
% ...
}

\begin{affiliations}
\item Astrophysical Institute, Vrije Universiteit Brussel, Pleinlaan 2, 1050 Brussels, Belgium 
\item Department of Astrophysics/IMAPP, Radboud University Nijmegen, P.O. Box 9010, 6500 GL Nijmegen, The Netherlands 
\item ASTRON, Netherlands Institute for Radio Astronomy, Postbus 2, 7990 AA, Dwingeloo, The Netherlands 
\item Nikhef, Science Park Amsterdam, 1098 XG Amsterdam, The Netherlands 
\item Max-Planck-Institut f\"{u}r Radioastronomie, Auf dem H\"ugel 69, 53121 Bonn, Germany 
\item IKP, Karlsruhe Institute of Technology (KIT), Postfach 3640, 76021 Karlsruhe, Germany 
\item Department of Physics and Astronomy, University of California Irvine, Irvine, CA 92697, USA 
\item KVI CART, University of Groningen, 9747 AA Groningen, The Netherlands 
\item Vrije Universiteit Brussel, Dienst ELEM, Brussels, Belgium 
\item Helmholtz-Zentrum Potsdam, DeutschesGeoForschungsZentrum GFZ, Department 1: Geodesy and Remote Sensing, Telegrafenberg, A17, 14473 Potsdam, Germany 
\item Shell Technology Center, Bangalore, India 
\item SRON Netherlands Insitute for Space Research, PO Box 800, 9700 AV Groningen, The Netherlands 
\item Kapteyn Astronomical Institute, PO Box 800, 9700 AV Groningen, The Netherlands 
\item CSIRO Australia Telescope National Facility, PO Box 76, Epping NSW 1710, Australia 
\item University of Twente, The Netherlands 
\item Harvard-Smithsonian Center for Astrophysics, 60 Garden Street, Cambridge, MA 02138, USA 
\item SKA South Africa, 3rd Floor, The Park, Park Road, Pinelands, 7405, South Africa 
\item Institute for Astronomy, University of Edinburgh, Royal Observatory of Edinburgh, Blackford Hill, Edinburgh EH9 3HJ, UK 
\item University of Hamburg, Gojenbergsweg 112, 21029 Hamburg, Germany 
\item Leibniz-Institut f\"{u}r Astrophysik Potsdam (AIP), An der Sternwarte 16, 14482 Potsdam, Germany 
\item School of Physics and Astronomy, University of Southampton, Southampton, SO17 1BJ, UK 
\item Research School of Astronomy and Astrophysics, Australian National University, Canberra, ACT 2611 Australia 
\item Anton Pannekoek Institute for Astronomy, University of Amsterdam, Science Park 904, 1098 XH Amsterdam, The Netherlands 
\item Max Planck Institute for Astrophysics, Karl Schwarzschild Str. 1, 85741 Garching, Germany 
\item Onsala Space Observatory, Dept. of Earth and Space Sciences, Chalmers University of Technology, SE-43992 Onsala, Sweden 
\item SmarterVision BV, Oostersingel 5, 9401 JX Assen 
\item Astronomisches Institut der Ruhr-Universit\"{a}t Bochum, Universitaetsstrasse 150, 44780 Bochum, Germany 
\item Th\"{u}ringer Landessternwarte, Sternwarte 5, D-07778 Tautenburg, Germany 
\item Hamburger Sternwarte, Gojenbergsweg 112, D-21029 Hamburg 
\item Laboratoire Lagrange, Universit\'e C\^ote d'Azur, Observatoire de la C\^ote d'Azur, CNRS, Blvd de l'Observatoire, CS 34229, 06304 Nice cedex 4, France 
\item Leiden Observatory, Leiden University, PO Box 9513, 2300 RA Leiden, The Netherlands 
\item LPC2E - Universite d'Orleans/CNRS 
\item Station de Radioastronomie de Nancay, Observatoire de Paris - CNRS/INSU, USR 704 - Univ. Orleans, OSUC , route de Souesmes, 18330 Nancay, France 
%\item 87801-0387, USA
\item National Radio Astronomy Observatory, 1003 Lopezville Road, Socorro, NM 87801-0387, USA 
\item Astrophysics, University of Oxford, Denys Wilkinson Building, Keble Road, Oxford OX1 3RH 
\item Astro Space Center of the Lebedev Physical Institute, Profsoyuznaya str. 84/32, Moscow 117997, Russia 
\item National Astronomical Observatory of Japan, Japan 
\item Sodankyl\"{a} Geophysical Observatory, University of Oulu, T\"{a}htel\"{a}ntie 62, 99600 Sodankyl\"{a}, Finland 
\item STFC Rutherford Appleton Laboratory,  Harwell Science and Innovation Campus,  Didcot  OX11 0QX, UK 
\item Center for Information Technology (CIT), University of Groningen, The Netherlands 
\item Centre de Recherche Astrophysique de Lyon, Observatoire de Lyon, 9 av Charles Andr\'{e}, 69561 Saint Genis Laval Cedex, France 
\item Fakult\"{a}t f\"uŸr Physik, Universit\"{a}t Bielefeld, Postfach 100131, D-33501, Bielefeld, Germany 
\item Department of Physics and Electronics, Rhodes University, PO Box 94, Grahamstown 6140, South Africa 
\item Jodrell Bank Center for Astrophysics, School of Physics and Astronomy, The University of Manchester, Manchester M13 9PL,UK 
\item Department of Astrophysical Sciences, Princeton University, Princeton, NJ 08544, USA 
\item GEPI, Observatoire de Paris, CNRS, Universit\'e Paris Diderot, 5 place Jules Janssen, 92190 Meudon, France
\item LESIA, UMR CNRS 8109, Observatoire de Paris, 92195 Meudon, France

\end{affiliations}

\begin{abstract}
Cosmic rays are the highest energy particles found in nature.
Measurements of the mass composition of cosmic rays between $10^{17}$ eV and $10^{18}$ eV are essential to understand whether this energy range is dominated by Galactic or extragalactic sources. 
It has also been proposed that the astrophysical neutrino signal\cite{icecube_science} comes from accelerators capable of producing cosmic rays of these energies\cite{MAL}.
Cosmic rays initiate cascades of secondary particles (air showers) in the atmosphere and their masses are inferred from measurements of the atmospheric depth of the shower maximum, $X_{\mathrm{max}}$\cite{augercomp}, or the composition of shower particles reaching the ground\cite{KG-ankle}. Current measurements\cite{KU12} suffer from either low precision, 
and/or a low duty cycle. 
%and a high energy threshold. 
Radio detection of cosmic rays\cite{Allen,FG03, LOPES05} is a rapidly developing technique\cite{HuegeICRC}, suitable for determination of $X_{\mathrm{max}}$\cite{lopes_ldf, anita_belov} with a duty cycle of in principle nearly 100\%. The radiation is generated by the separation of relativistic charged particles in the geomagnetic field and a negative charge excess in the shower front\cite{Allen, scholten}.
Here we report radio measurements of $X_{\mathrm{max}}$ with a mean precision of 16~g/cm$^2$ between $10^{17}-10^{17.5}$ eV.
Because of the high resolution in $X_{\mathrm{max}}$ we can determine the mass spectrum and find a mixed composition, containing a light mass fraction of $\sim80$\%. 
Unless the extragalactic component becomes significant already below $10^{17.5}$ eV, our measurements indicate an additional Galactic component dominating at this energy range.
\end{abstract}

Observations were made with the Low Frequency Array (LOFAR\cite{bib:LOFAR}), a radio telescope consisting of thousands of crossed dipoles, with built-in air shower detection capability\cite{pipeline}. LOFAR records the radio signals from air showers continuously while running astronomical observations simultaneously.
It comprises a scintillator array (LORA), that triggers the readout of buffers, storing the full waveforms received by all antennas.

We have selected air showers from the period June 2011 - January 2015 with radio pulses in at least 192 antennas. 
The total uptime was $\sim$150 days, limited by construction and commissioning of the telescope. 
Showers that occurred within an hour from lightning activity, or have a polarisation pattern that is indicative of influences from atmospheric electric fields are excluded from the sample\cite{polprl}. 

Radio intensity patterns from air showers are asymmetric due to the interference between geomagnetic and charge excess radiation. They can be reproduced from first principles by summing the radio contributions of all electrons and positrons in the shower. We use the radio simulation code CoREAS\cite{coreas}, a plug-in of CORSIKA\cite{corsika}, which follows this approach. 

It has been shown that $X_{\mathrm{max}}$ can be accurately reconstructed from densely sampled radio measurements\cite{Buitink2014}. We use a hybrid approach, simultaneously fitting the radio and  particle data. The radio component is very sensitive to $X_{\mathrm{max}}$, while the particle component is used for the energy measurement.

\begin{figure}[H]
\centering
\includegraphics[width=0.8\textwidth]{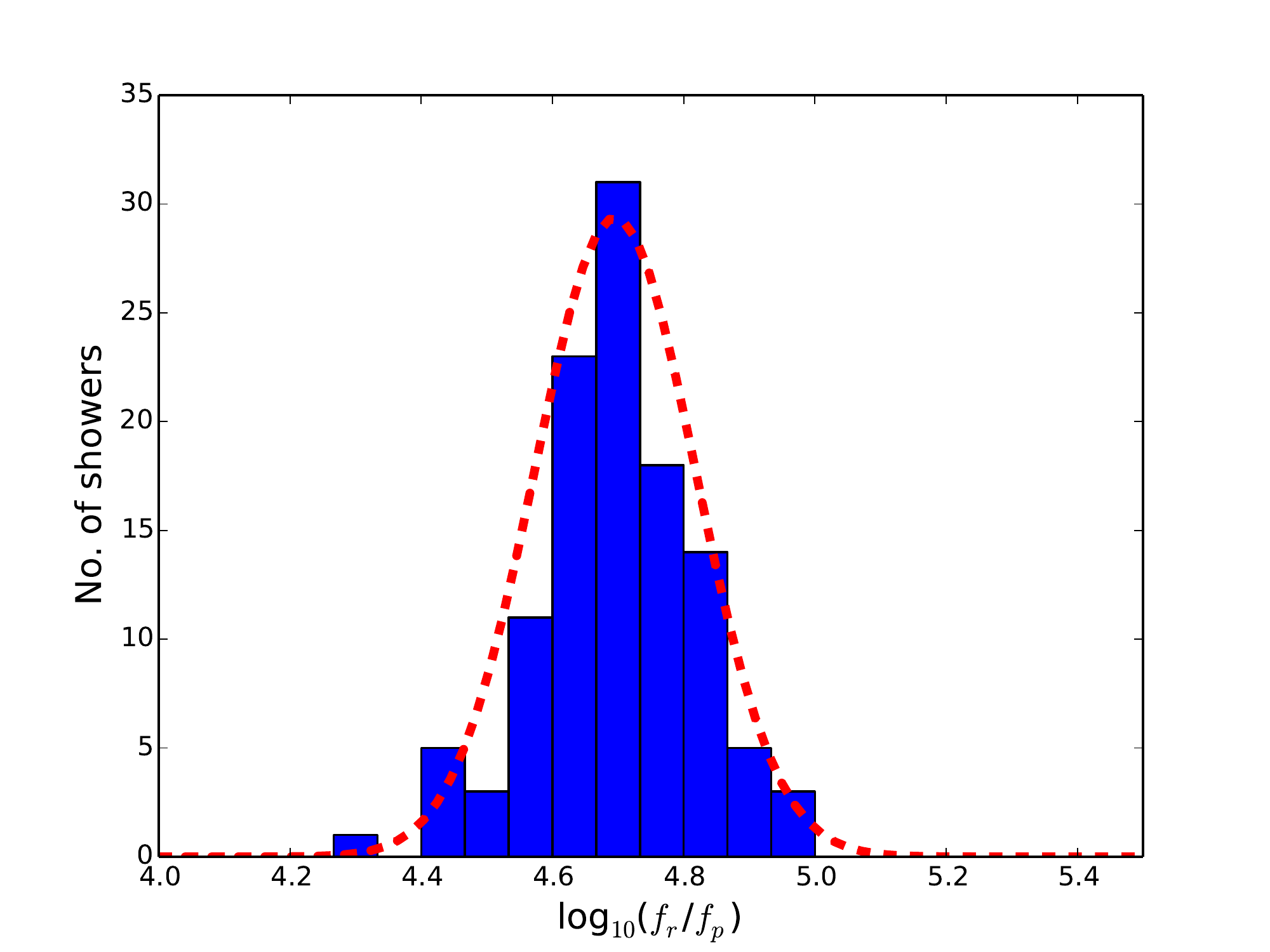}
\caption{{\bf Figure 1 $|$ Energy resolution.} The distribution of $f_r/f_p$ is fitted with a Gaussian, yielding $\sigma=0.12$ on a logarithmic scale, corresponding to an energy resolution of 32\%. This value is actually the quadratic sum of the energy resolution of the radio and particle resolutions. In this analysis, there was no absolute calibration for the received radio power yet, so $f_r$ has an arbitrary scale.}
\label{fig:resolution}
\end{figure}

The fit contains four free parameters: the shower core position $(x,y)$, and scaling factors for the particle density $f_p$ and the radio power $f_r$. If $f_p$ deviates significantly from unity, the reconstructed energy does not match the simulation and a new set of simulations is produced. This procedure is repeated until the energies agree within uncertainties.
The ratio between $f_r$ and $f_p$ should be the same for all showers and is used to derive the energy resolution of 32\% (see Figure \ref{fig:resolution}). 

The radio intensity fits have reduced $\chi^2$-values ranging from 0.9 to 2.9. All features in the data are well reproduced by the simulation (see Extended Data Figs.~1-5), demonstrating that the radiation mechanism is now well-understood. The reduced $\chi^2$-values exceeding unity may indicate remaining uncertainties in the antenna response, atmospheric properties, or limitations of the simulation software. 

Radio detection becomes more efficient for higher-altitude showers that have larger footprints. The particle trigger, however, becomes less efficient since the number of particles reaching the ground decreases. To avoid a bias, we require that \emph{all} the simulations produced for a shower pass the trigger criteria. Above $10^{17}$~eV this cut removes 4 showers from the sample. At lower energies, this number rapidly increases, and we exclude all showers below $10^{17}$~eV from this analysis. 

Furthermore, we evaluate the reconstructed core positions of all simulated showers. Showers with a mean reconstruction error above 5~m are rejected. This cut does not introduce a composition bias, because it is based on the sets of simulated showers, and not on the data. The final event sample contains 118 showers.

\begin{figure}[H]
\centering
\includegraphics[width=0.8\textwidth]{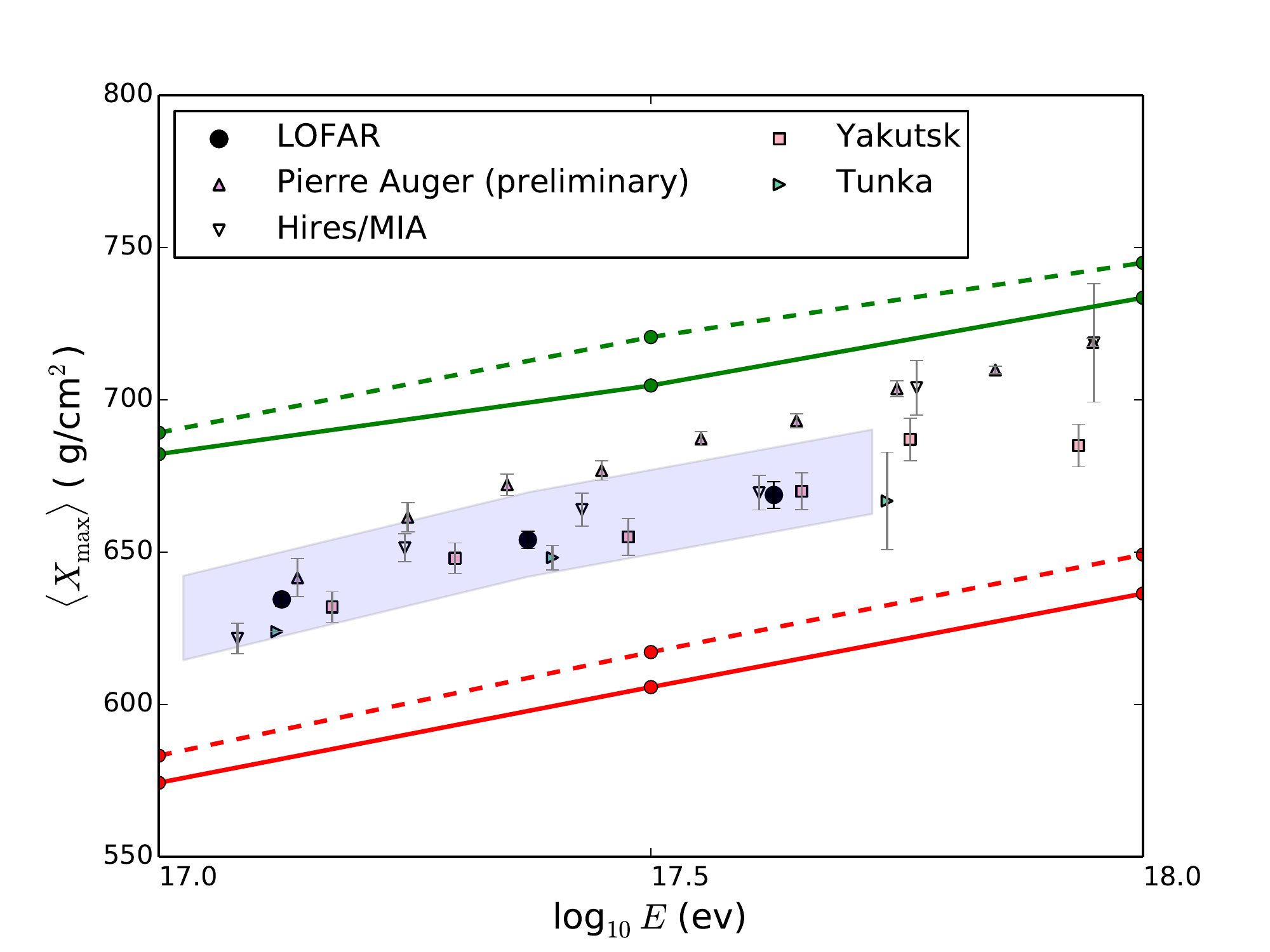}
\caption{{\bf Figure 2 $|$ Measurements of $\langle X_{\mathrm{max}} \rangle$}. The mean depth of shower maximum as a function of energy is plotted for LOFAR and earlier experiments based on different techniques\cite{escale,hiresmia,tunka,yakutsk}. Error bars indicate one-sigma uncertainties. The systematic uncertainties are +14/-10 g/cm$^2$ on $\langle X_{\mathrm{max}}\rangle$ and 27\% on energy and are indicated with a shaded band. The Pierre Auger Observatory measures the fluorescence light emitted by atmospheric molecules excited by air shower particles. Hires/MIA used a combination of the fluorescence technique and muon detection. The Tunka and Yakutsk arrays use non-imaging Cherenkov detectors. The green (upper) lines indicate the  $\langle X_{\mathrm{max}}\rangle$ for proton shower simulated with QGSJETII.04 (solid) and EPOS-LHC (dashed). The blue (lower) lines are for showers initiated by iron nuclei.}
\label{fig:xmax}
\end{figure}

The uncertainty on $X_{\mathrm{max}}$ is determined independently for all showers\cite{Buitink2014}, and has a mean value of 16 g/cm$^{2}$ (see Extended Data Figure~6). Figure~\ref{fig:xmax} shows our measurements of the {\it average} $X_{\mathrm{max}}$, which are consistent with earlier experiments using different methods, within statistical uncertainties.
The high resolution for $X_{\mathrm{max}}$ \emph{per shower} allows us to derive
more information on the composition of cosmic rays, by studying the complete shape of the $X_{\mathrm{max}}$ distribution. For each shower, we calculate:
\begin{equation}
a = \frac{\langle X_\mathrm{proton}\rangle - X_\mathrm{shower}}{\langle X_\mathrm{proton}\rangle - \langle X_\mathrm{iron}\rangle}
\label{eq}
\end{equation}   
where $X_\mathrm{shower}$ is the reconstructed $X_{\mathrm{max}}$, and $ \langle X_\mathrm{proton}\rangle$ and $ \langle X_\mathrm{iron}\rangle$ are mean values predicted by the hadronic interaction code QGSJETII.04\cite{qgsjet}. 

\begin{figure}[H]
\centering
\includegraphics[width=0.8\textwidth]{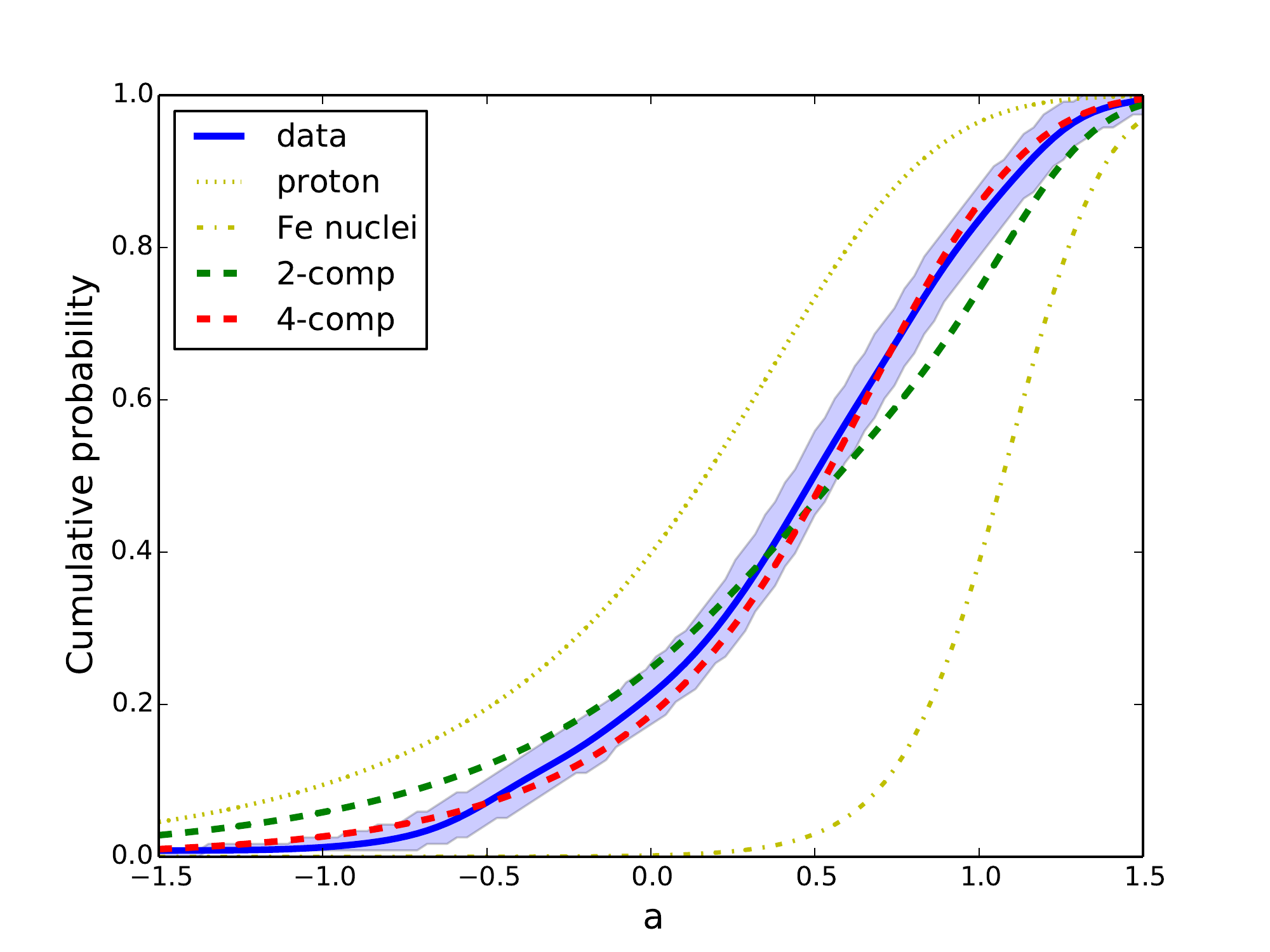}
\caption{{\bf Figure 3 $|$ Composition model fits.} The cumulative probability density of the parameter $a$ (see Eqn.~\ref{eq}) is plotted for the data (blue line, shading indicates the range where the p-value$>0.01$) and several models, based on QGSJETII.04 simulations.
A set that contains only proton showers is centered around $a=0$ and has a large spread, while iron showers give a small spread around $a=1$.  A two-component model of proton and iron yields the best fit for a proton fraction of 62\%, but does not describe the data well with a p-value of $1.1\times 10^{-6}$. A four-component model gives the best fit at 0\% proton, 79\% helium, 19\% nitrogen, and 2\% iron,with a p-value of $0.17$. The uncertainty on these values is explored in Figure~\ref{fig:contour}.} 
\label{fig:cdf}
\end{figure}

The cumulative probability density function (CDF) for all showers is plotted in Fig.~\ref{fig:cdf}. First, we fit a 2-component model of proton and iron nuclei, with the mixing ratio as the only free parameter. To calculate the corresponding CDFs we use a parametrisation of the $X_{\mathrm{max}}$ distribution fitted to simulations based on QGSJETII.04. 
The best fit is found for a proton fraction of 62\%, but it describes the data poorly with a p-value of $1.1\times 10^{-6}$.

\begin{figure}[H]
\centering
\includegraphics[width=0.8\textwidth]{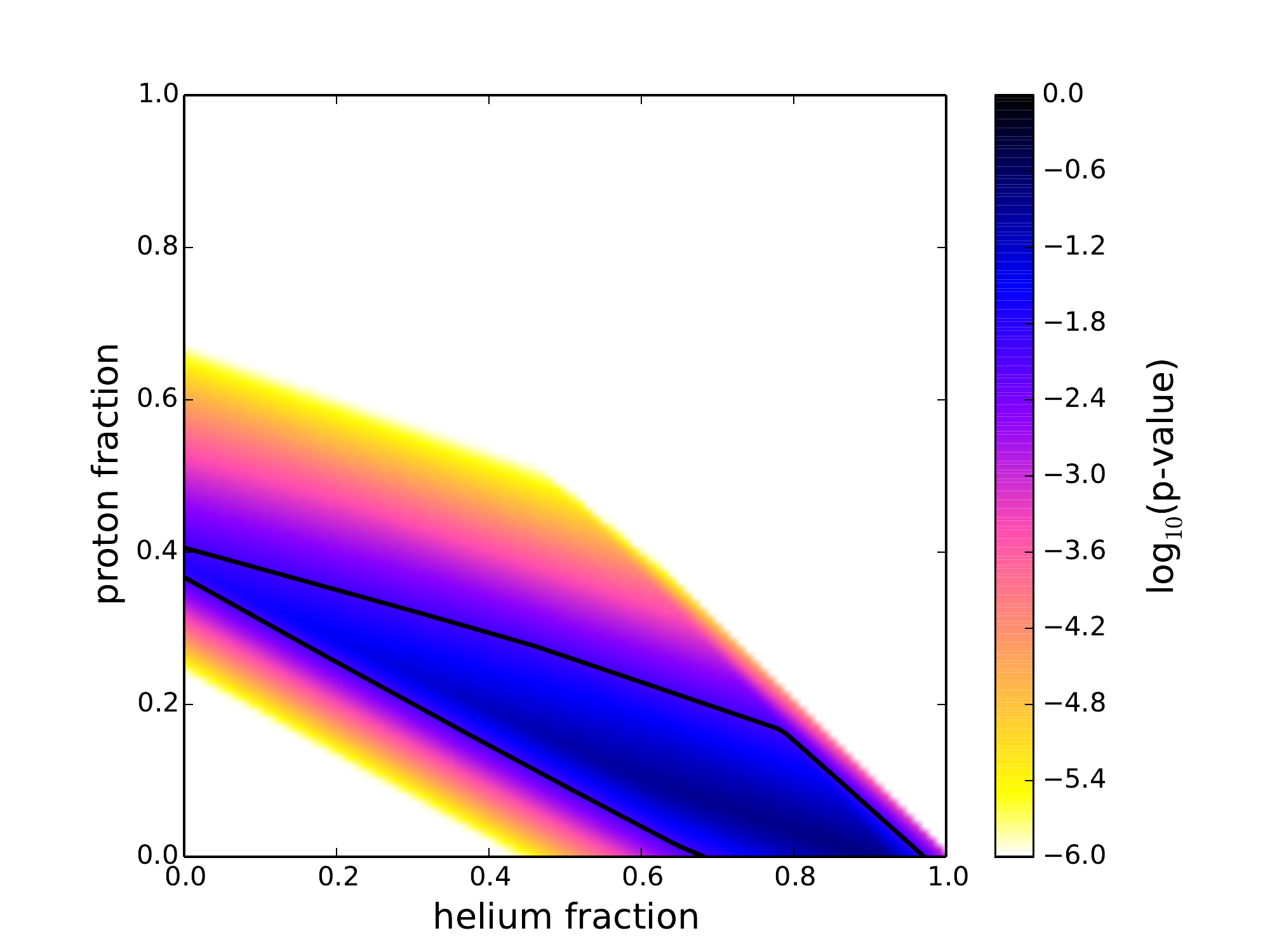}
\caption{{\bf Figure 4 $|$ p-value distribution for four-component model.} The four-component model is explored further by keeping the proton and helium fractions fixed at all possible combinations, and solving for the nitrogen/iron ratio. The p-value (see Fig.~\ref{fig:cdf}) is plotted as a function of the proton and helium fraction. The optimal fit is found at 0\% proton and 79\% helium (p=0.17), but the deviation only deteriorates slowly when replacing helium with proton. The thick black contour line contains all combinations for which $p>0.01$. At this significance level the total fraction of light elements (p+He) lies between 0.38 and 0.98.}
\label{fig:contour}
\end{figure}

A better fit is achieved with a four-component model (p+He+N+Fe), yielding a p-value of 0.17. While the best fit is found for a Helium fraction of 80\%, the fit quality deteriorates only slowly when replacing helium by protons. This is demonstrated in Figure~\ref{fig:contour} where the p-value is plotted for four-component fits where the fractions of helium and proton are fixed, and the ratio between N and Fe is left as the only free parameter. 
The total fraction of light elements (p+He) is in the range [0.38,0.98] at 99\% confidence level, with a best fit value of 0.8.  
The heaviest composition that is allowed within systematic uncertainties still has a best fit p+He fraction of 0.6, and a 99\% confidence level interval of [0.18, 0.82]. The online method section contains information about the systematic uncertainties and the statistical analysis.

The abundances of individual elements depend on the hadronic interaction model. The $X_{\mathrm{max}}$ values predicted by EPOS-LHC\cite{epos} are on average 15-20 g/cm$^2$ higher than QGSJETII.04 (see Fig.~\ref{fig:xmax}). This coincides with the separation between, for example, protons and deuterium or between helium and beryllium. We therefore prefer to present our result as a total fraction of light elements, instead of placing too much emphasis on individual elements.  

Recent results for the Pierre Auger Observatory indicate that the cosmic ray composition at 10$^{18}$ eV, just below the ankle, can be fitted with a mixture of protons and either helium (QGSJET.II04) or nitrogen (EPOS-LHC)\cite{augercomp}. With decreasing energy, their proton fraction drops, while their helium (or nitrogen) fraction rises, down to the threshold energy of $7\cdot 10^{17}$~eV. An extrapolation of this trend to our mean energy of $3\cdot 10^{17}$~eV connects smoothly to our best fitting solution in which helium dominates.

KASCADE-Grande has reported an ankle-like feature at $10^{17.1}$ eV, where the spectral index for light elements changes to $\gamma=-2.79\pm0.08$\cite{KG-ankle}. However, they find a light particle (p+He) fraction below 30\% at $3\times 10^{17}$~eV (based on their Figure 4), considerably lower than our value. In contrast to LOFAR and Auger, their composition measurements are based on the muon/electron ratio. Auger has reported a muon excess compared to all commonly used hadronic interaction models\cite{augermuons}. Inaccurate predictions of muon production, or $\langle X_{\mathrm{max}} \rangle$, can be the cause of the discrepancy in the fraction of light particles between LOFAR and KASCADE-Grande.  

If the knee in the all-particle spectrum near $3\times 10^{15}$~eV corresponds to the proton or helium cut-off of the main Galactic cosmic-ray population, the corresponding iron cut-off would lie at most at an energy 26 times larger. If this population still dominates at 10$^{17}$~eV, the mass composition should be dominated by heavy elements at that energy. Therefore, the large component of light elements observed with LOFAR must have another origin.

In principle, it is possible that we observe an extragalactic component.
 In that case the ankle in the cosmic-ray spectrum, slightly above 10$^{18}$ eV, does not indicate the transition from Galactic to extragalactic origin. Instead, it can be explained as the imprint of pair production on the cosmic microwave background on an extragalactic proton spectrum\cite{aloisio}. However, since this feature only appears for a proton-dominated flux it is in tension with our data that favours a mixture of light elements.

A second Galactic component, dominating around 10$^{17}$~eV, can be produced by a class of extremely energetic sources (Galactic exatrons), like the explosions of Wolf Rayet stars into their stellar winds\cite{SBG}, or past Galactic gamma-ray bursts\cite{calvez2010}. 
Alternatively, the original Galactic population could be reaccelerated by the Galactic wind termination shock\cite{JM87}. Such scenarios predict mixtures of light elements, consistent with our results.

\begin{oldthebibliography}{}
\bibitem{icecube_science} Aartsen, M. et al. [IceCube collaboration], Evidence for High-Energy Extraterrestrial Neutrinos at the IceCube Detector, {\it Science} {\bf 342}, 1242856 (2013).
\bibitem{MAL}Murase, K., Ahlers, M., and Lacki, B., Testing the hadronuclear origin of PeV neutrinos observed with IceCube, {\it Phys. Rev. D} {\bf 88}, 121301 (2013).
\bibitem{augercomp} Aab, A. et al. [Pierre Auger collaboration], Depth of maximum of air-shower profiles at the Pierre Auger Observatory. II. Composition implications, {\it Phys. Rev. D} {\bf 90}, 122006 (2014).
\bibitem{KG-ankle} Apel, W. et al. [KASCADE-Grande collaboration], Ankle-like feature in the energy spectrum of light elements of cosmic rays observed with KASCADE-Grande, {\it Phys. Rev. D} {\bf 87}, 081101 (2013).
\bibitem{KU12}Kampert, K.-H. and Unger, M., Measurements of the cosmic ray composition with air shower experiments, {\it Astropart. Phys.} {\bf 35}, 660-678 (2012).
\bibitem{Allen} Allan, H. R., Radio Emission from Extensive Air Showers, {\it Prog. Elem. Part. Cosm. Ray Phys.} {\bf 10}, 171-302 (1971).
\bibitem{FG03} Falcke, H. \& Gorham, P.W., Detecting radio emission from cosmic ray air showers and neutrinos with a digital radio telescope, {\it Astropart. Phys.} {\bf 19}, 477-494 (2003).
\bibitem{LOPES05} Falcke, H. et al. [LOPES collaboration], Detection and imaging of atmospheric radio flashes from cosmic ray air showers, {\it Nature} {\bf 435}, 313-316 (2005).
\bibitem{HuegeICRC} Huege, T., The Renaissance of Radio Detection of Cosmic Rays, {\it Brazilian Journal of Physics} {\bf 44}, 520-529 (2014). 
\bibitem{lopes_ldf} Apel, W. et al. [LOPES collaboration], Reconstruction of the energy and depth of maximum of cosmic-ray air showers from LOPES radio measurements, {\it Phys. Rev. D} {\bf 90}, 062001 (2014).
\bibitem{anita_belov}Belov, K. et al. [ANITA collaboration], Towards determining the energy of the UHECRs observed by the ANITA detector, {\it AIP Conference Proceedings} {\bf 1535}, 209-213 (2013).
\bibitem{scholten}Werner, K. and Scholten, O., Macroscopic Treatment of Radio Emission from Cosmic Ray Air Showers based on Shower Simulations, {\it Astropart. Phys.} {\bf 29}, 393-411 (2008).  
\bibitem{bib:LOFAR} Van Haarlem, M. et al. [LOFAR collaboration], LOFAR: The LOw-Frequency ARray, {\it Astron. Astrophys.} {\bf  556}, A2 (2013).
\bibitem{pipeline} Schellart, P. et al. [LOFAR collaboration], Detecting cosmic rays with the LOFAR radio telescope, {\it Astron. Astrophys.} {\bf 560} A98 (2013).
\bibitem{polprl} Schellart, P. et al. [LOFAR collaboration], Probing Atmospheric Electric Fields in Thunderstorms through Radio Emission from Cosmic-Ray-Induced Air Showers, {\it Phys. Rev. Lett.} {\bf 114}, 165001 (2015).
\bibitem{coreas} Huege, T., Ludwig, M. and James,C., Simulating radio emission from air showers with CoREAS, {\it AIP Conference Proceedings} {\bf 1535}, 128-132 (2012).
\bibitem{corsika} Heck, D. et al., CORSIKA: a Monte Carlo code to simulate extensive air showers, {\it Report FZKA} {\bf 6019} (1998).
\bibitem{Buitink2014} Buitink, S. et al. [LOFAR collaboration], Method for high precision reconstruction of air shower X$_{max}$ using two-dimensional radio intensity profiles, {\it Phys.\ Rev.\ D} {\bf 90}, 082003 (2014).
\bibitem{qgsjet} Ostapchenko, S., QGSJET-II: results for extensive air showers, {\it Nucl. Phys. B Proc. Suppl.} {\bf 151}, 147-150 (2006).
\bibitem{epos} Pierog, T., and Werner, K., EPOS Model and Ultra High Energy Cosmic Rays, {\it Nucl. Phys. B Proc. Suppl.} {\bf 196}, 102-105 (2009).
\bibitem{augermuons} Aab, A. et al. [Pierre Auger collaboration], Muons in air showers at the Pierre Auger Observatory: Mean number in highly inclined events, {\it Phys. Rev. D} {\bf 91}, 032003 (2015).
\bibitem{aloisio} Aloisio, R. et al., A dip in the UHECR spectrum and the transition from galactic to extragalactic cosmic rays, {\it Astropart. Phys.} {\bf 27}, 76-91 (2007).
\bibitem{SBG} Stanev, T., Biermann, P., and Gaisser, T., Cosmic rays. IV. The spectrum and chemical composition above 10 GeV, {\it Astron. Astrophys.} {\bf 274}, 902-915 (1993).
\bibitem{calvez2010} Calvez, A., Kusenko, S., and Nagataki, S., Role of Galactic sources and magnetic fields in forming the observed energy-dependent composition of ultrahigh-energy cosmic rays, {\it Phys. Rev. Lett.} {\bf 105}, 091101 (2010).
\bibitem{JM87} Jokipii, J.R. and Morfill, G., Ultra-high-energy cosmic rays in a galactic wind and its termination shock, {\it Astrophys. J.} {\bf 312}, 170-177 (1987).
%\bibitem{escale} Letessier-Selvon, A. et al. [Pierre Auger collaboration], Highlights from the Pierre Auger Observatory, {\it Proc. 33rd ICRC Rio de Janeiro} 1277 (2013).
%\bibitem{escale} Aab, A. et al. [Pierre Auger collaboration], The Pierre Auger Observatory: Contributions to the 34th International Cosmic Ray Conference (ICRC 2015), {\it Proc. 34th ICRC The Hague} (2015).
\bibitem{escale} Porcelli, A. for the Pierre Auger Collaboration, PoS (ICRC2015) 420; arXiv:1509.03732
\bibitem{hiresmia} Abu-Zayyad, T. et al. [HiRes/MIA collaboration], Measurement of the Cosmic-Ray Energy Spectrum and Composition from 10$^{17}$ to 10$^{18.3}$ eV Using a Hybrid Technique, {\it Astrophys. J.} {\bf 557} (2001) 686-699.
%\bibitem{hirescomp} Abbasi, R.U. et al. [HiRes collaboration], Indications of Proton-Dominated Cosmic-Ray Composition above 1.6 EeV, {\it Phys. Rev. Lett.} {\bf 104}, 161101 (2010).
\bibitem{yakutsk} Knurenko, S. and Sabourov, A. [Yakutsk collaboration], The depth of maximum shower development and its fluctuations: cosmic ray mass composition at $E_0 \geq 10^{17} eV$, {\it Proc. XVI ISVHECRI } (2010).
\bibitem{tunka} Berezhnev, S.F. et al. [Tunka collaboration.], Tunka-133: Primary Cosmic Ray Mass Composition in the Energy Range $6\times 10^{15} - 10^{18}$ eV, {\it Proc. 32nd ICRC Beijing} 209 (2011).
\end{oldthebibliography}

\begin{addendum}

\item We acknowledge financial support from  the Netherlands Organization for Scientific Research (NWO), VENI grant 639-041-130, the Netherlands Research School for Astronomy (NOVA), the Samenwerkingsverband Noord-Nederland (SNN) and the Foundation for Fundamental Research on Matter (FOM). We acknowledge funding from the European Research Council under the European Union's Seventh Framework Program
(FP/2007-2013) / ERC (grant agreement n.~227610) and under the European Union's Horizon 2020 research and innovation programme (grant agreement n.~640130).
LOFAR, the Low Frequency Array designed and constructed by ASTRON, has facilities in several countries, that are owned by various parties (each with their own funding sources), and that are collectively operated by the International LOFAR Telescope (ILT) foundation under a joint scientific policy. 

\item[Author Contributions]
All authors are part of the LOFAR collaboration and have contributed to the design, construction, calibration, and maintenance of LOFAR and/or LORA. The first thirteen authors constitute the Cosmic Ray Key Science Project and have contributed to the acquisition, calibration, and analysis of cosmic ray radio data and LORA data. The manuscript was written by S.B. and subjected to an internal collaboration-wide review process. All authors approved the final version of the manuscript.

\item[Author Information]Reprints and permissions information is available at www.nature.com/reprints. The authors declare no competing financial interests. Correspondence and requests for materials should be addressed to Stijn Buitink (\texttt{Stijn.Buitink@vub.ac.be}).
\end{addendum}

\newpage

\begin{figure}[H]
\centering
\includegraphics[clip=true,trim=65 0 65 15, width=0.9\textwidth]{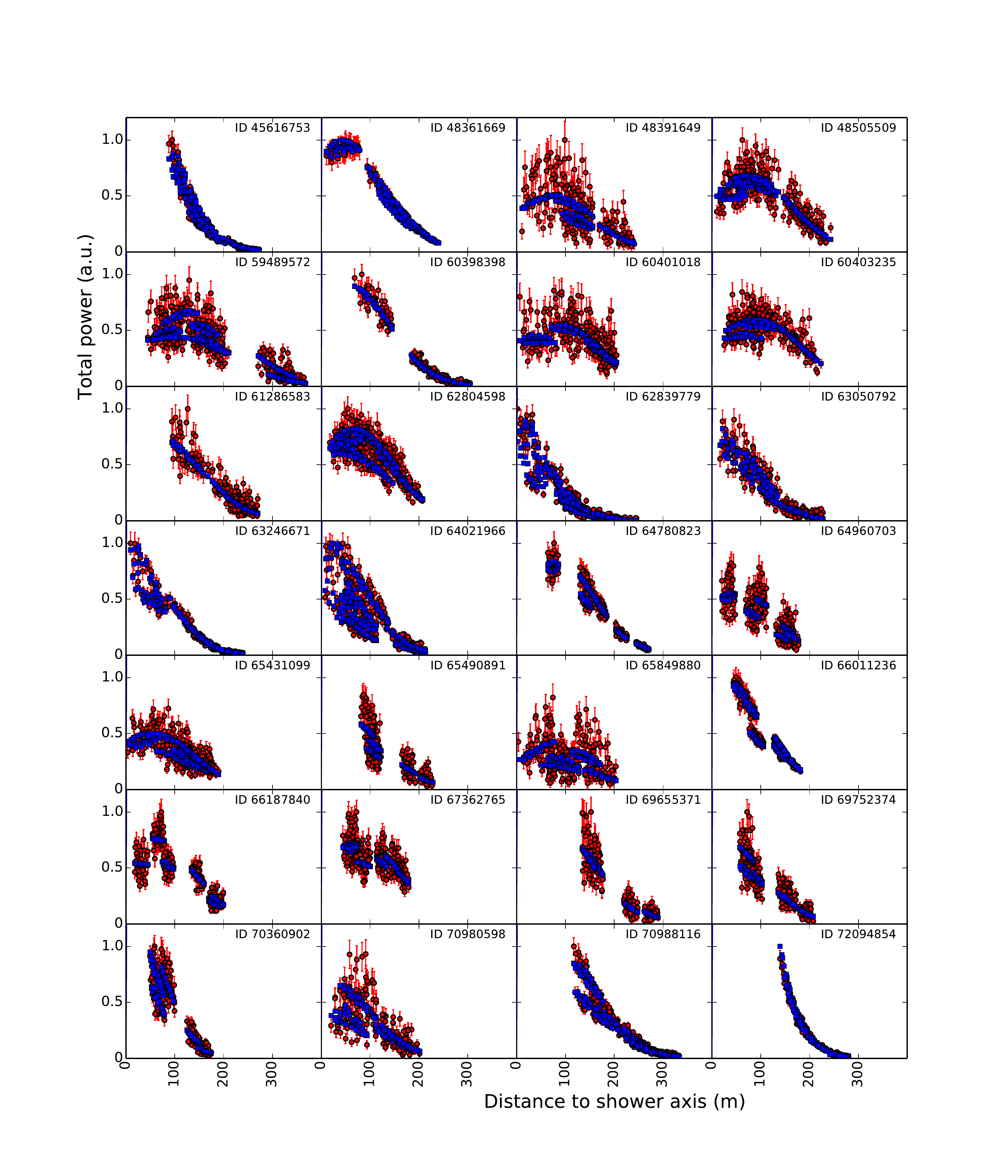}
\caption{{\bf Extended Data Figure 1 $|$ Fitted lateral distributions.} Lateral distribution of radio pulse power for all 118 measured showers (red circles) and the corresponding best-fitting CoREAS simulation (blue squares).}
\label{fig:list1}
\end{figure}

\begin{figure}[H]
\centering
\includegraphics[clip=true,trim=65 0 65 15, width=0.9\textwidth]{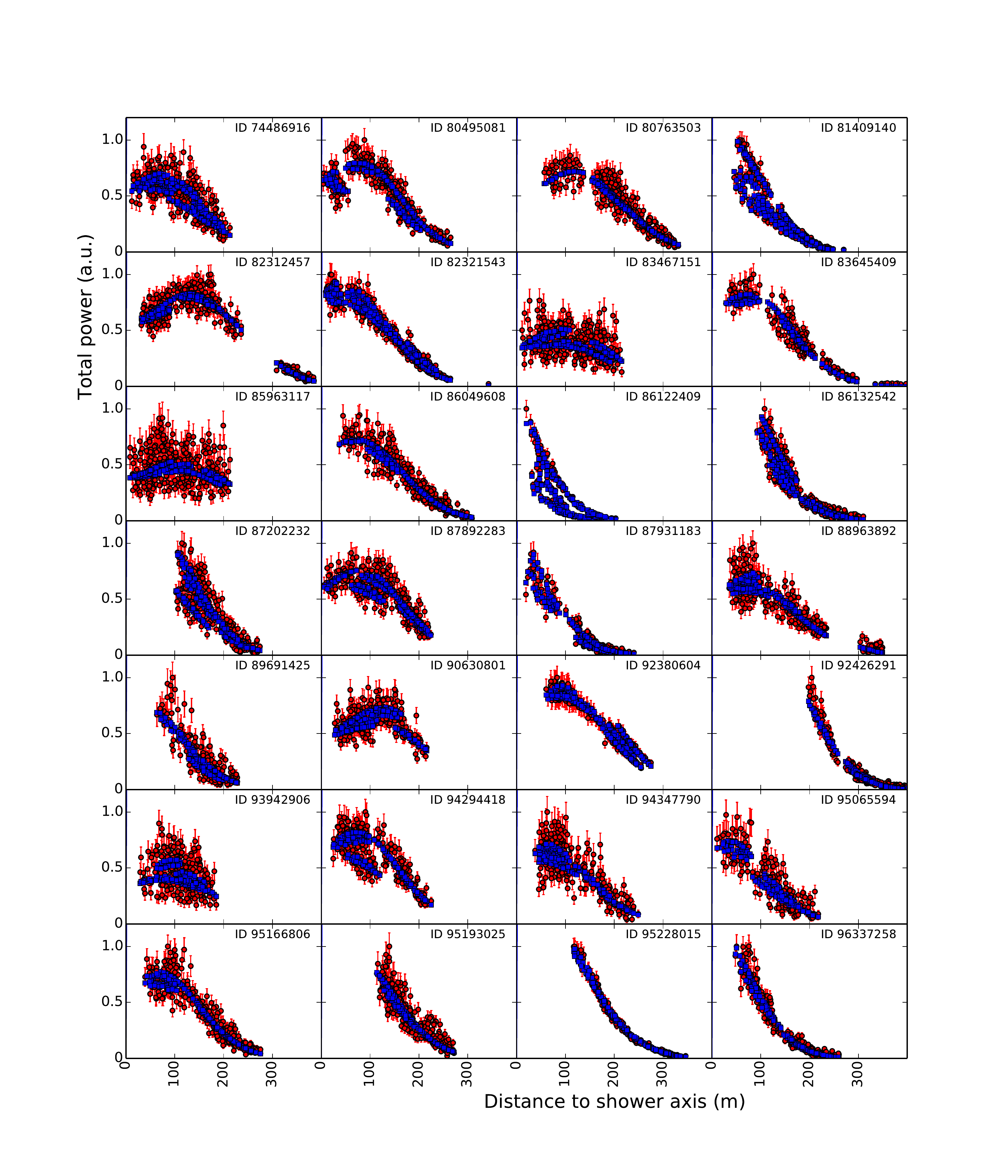}
\caption{{\bf Extended Data Figure 2 $|$ Fitted lateral distributions.} Continuation of Extended Data Figure 1.}
\end{figure}

\begin{figure}[H]
\centering
\includegraphics[clip=true,trim=65 0 65 15, width=0.9\textwidth]{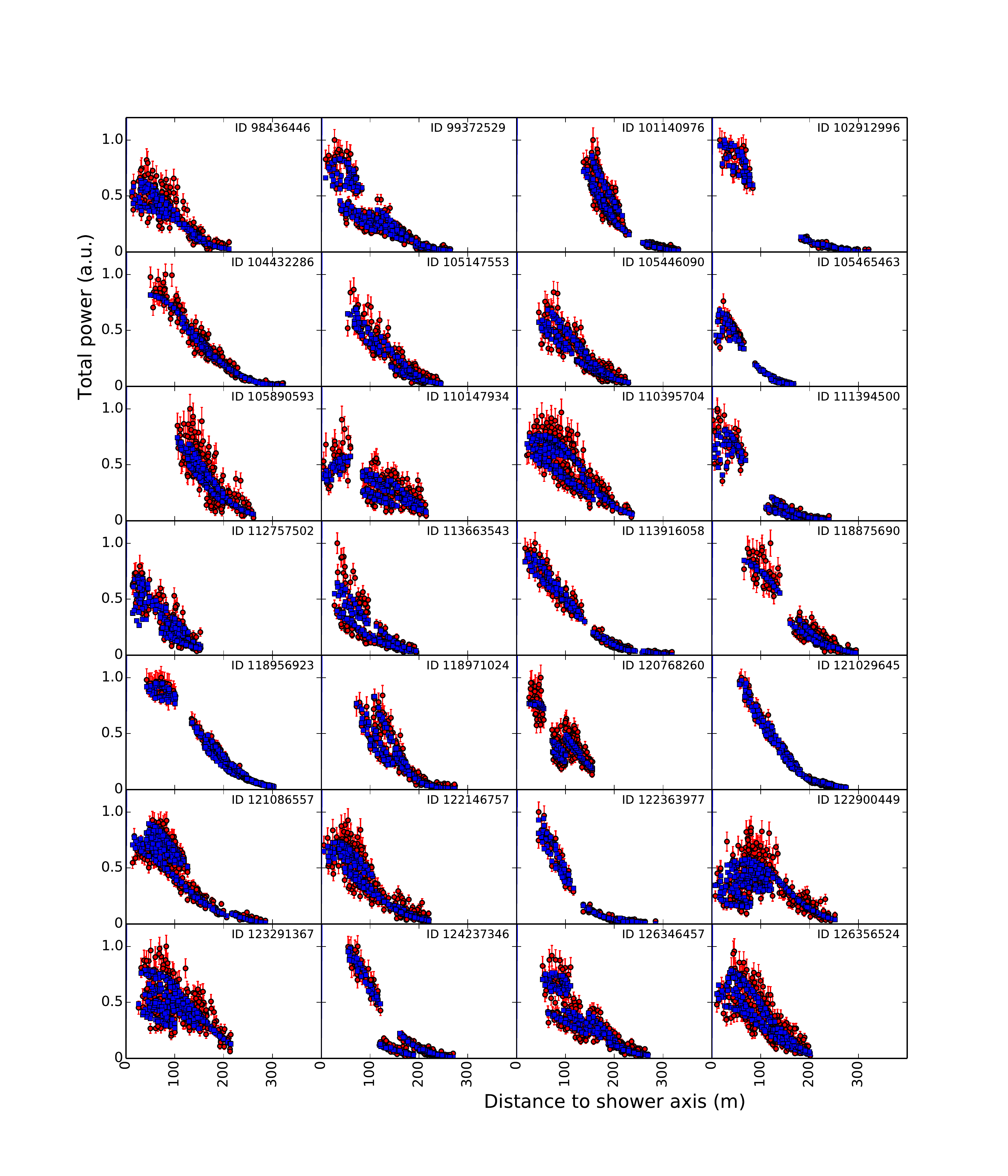}
\caption{{\bf Extended Data Figure 3 $|$ Fitted lateral distributions.} Continuation of Extended Data Figure 2.}
\end{figure}

\begin{figure}[H]
\centering
\includegraphics[clip=true,trim=65 0 65 15, width=0.9\textwidth]{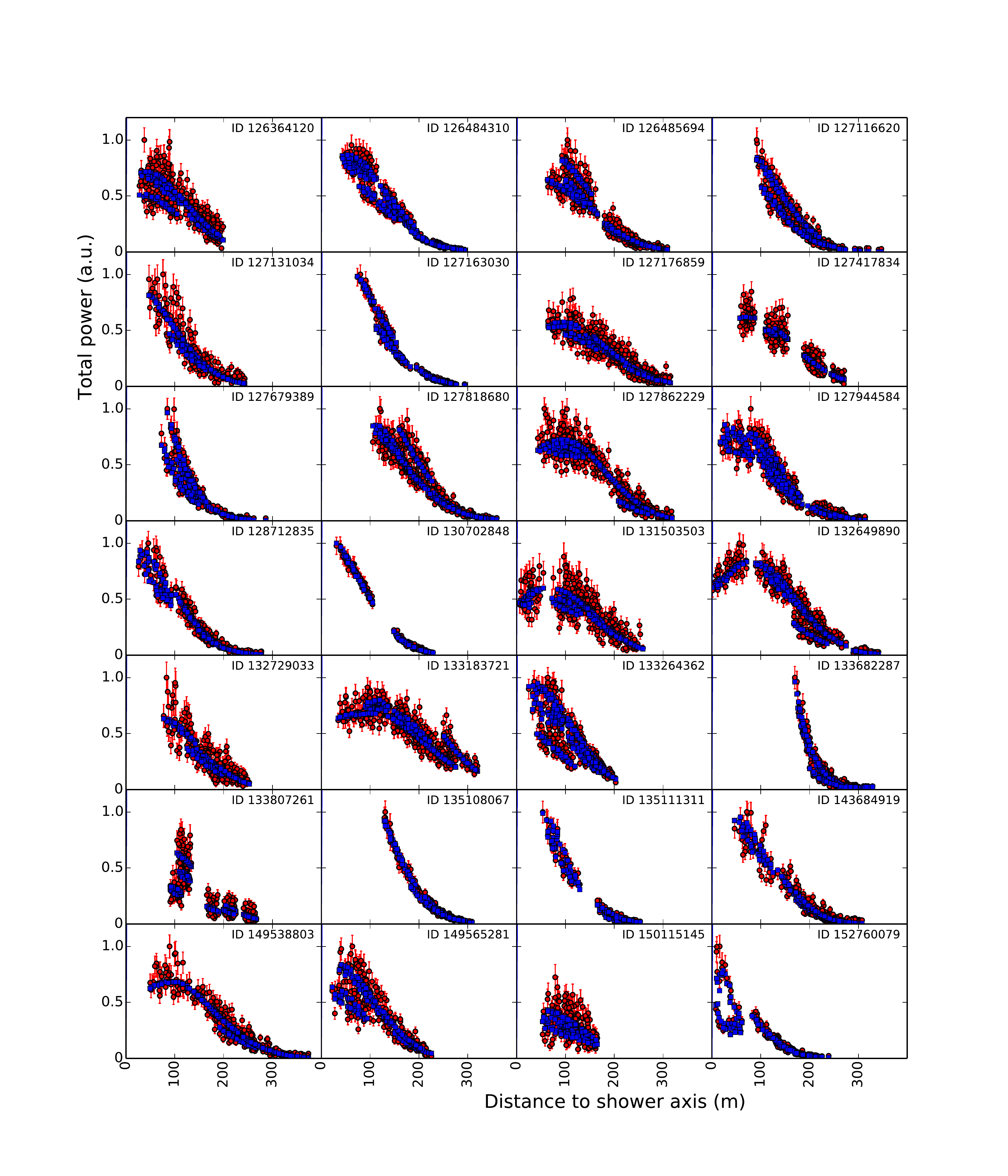}
\caption{{\bf Extended Data Figure 4 $|$ Fitted lateral distributions.} Continuation of Extended Data Figure 3.}
\end{figure}

\begin{figure}[H]
\centering
\includegraphics[width=0.9\textwidth]{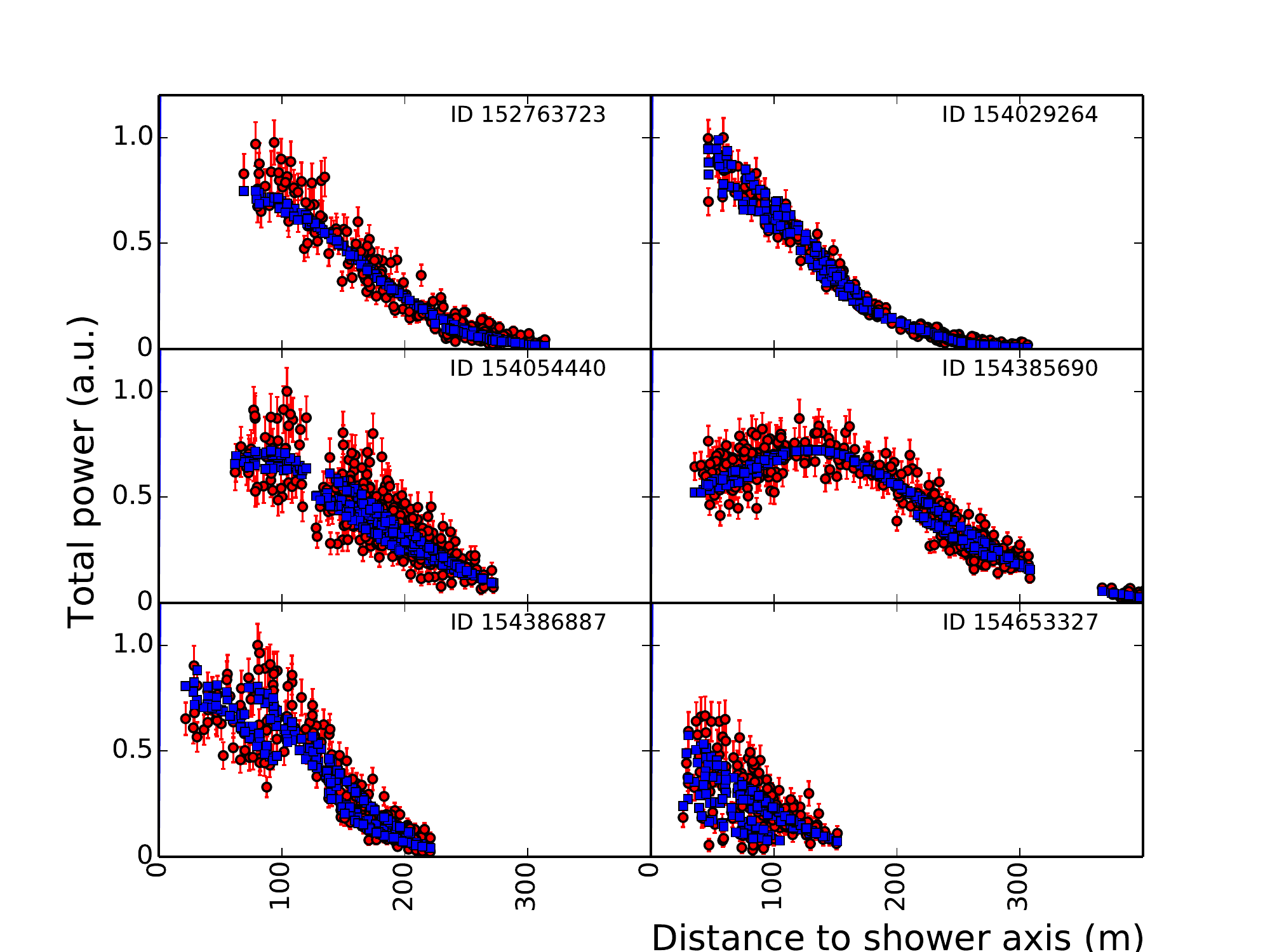}
\caption{{\bf Extended Data Figure 5 $|$ Fitted lateral distributions.} Continuation of Extended Data Figure 4.}
\end{figure}

\begin{figure}[H]
\centering
\includegraphics[width=0.8\textwidth]{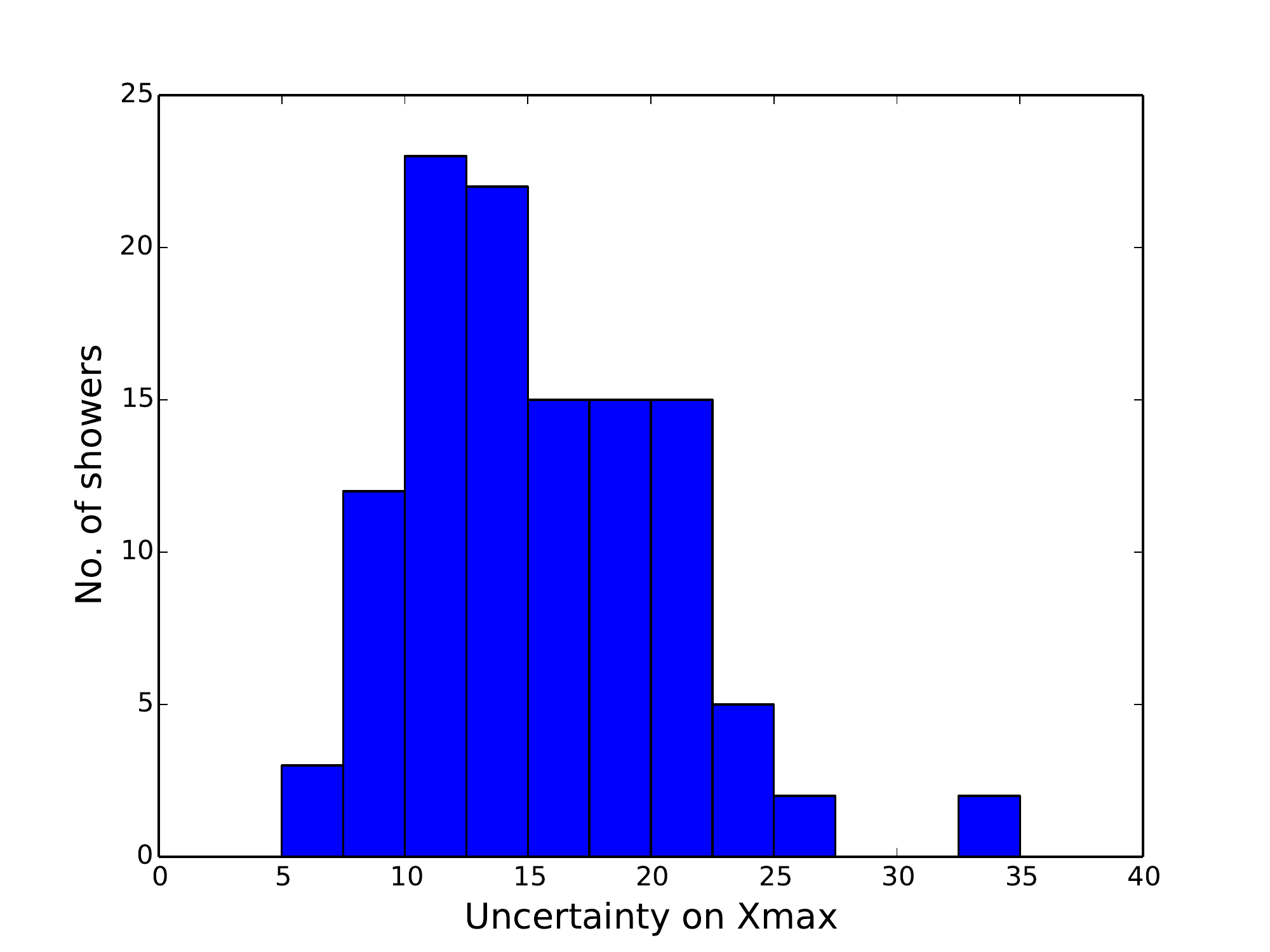}
\caption{{\bf Extended Data Figure 6 $|$ Distribution of uncertainty on  $X_{\mathrm{max}}$} The distribution of the uncertainty on $X_{\mathrm{max}}$ for all showers used in this analysis. The mean value is 16 g/cm$^2$.}
\label{fig:xerrdist}
\end{figure}

\begin{figure}[H]
\centering
\includegraphics[width=0.8\textwidth]{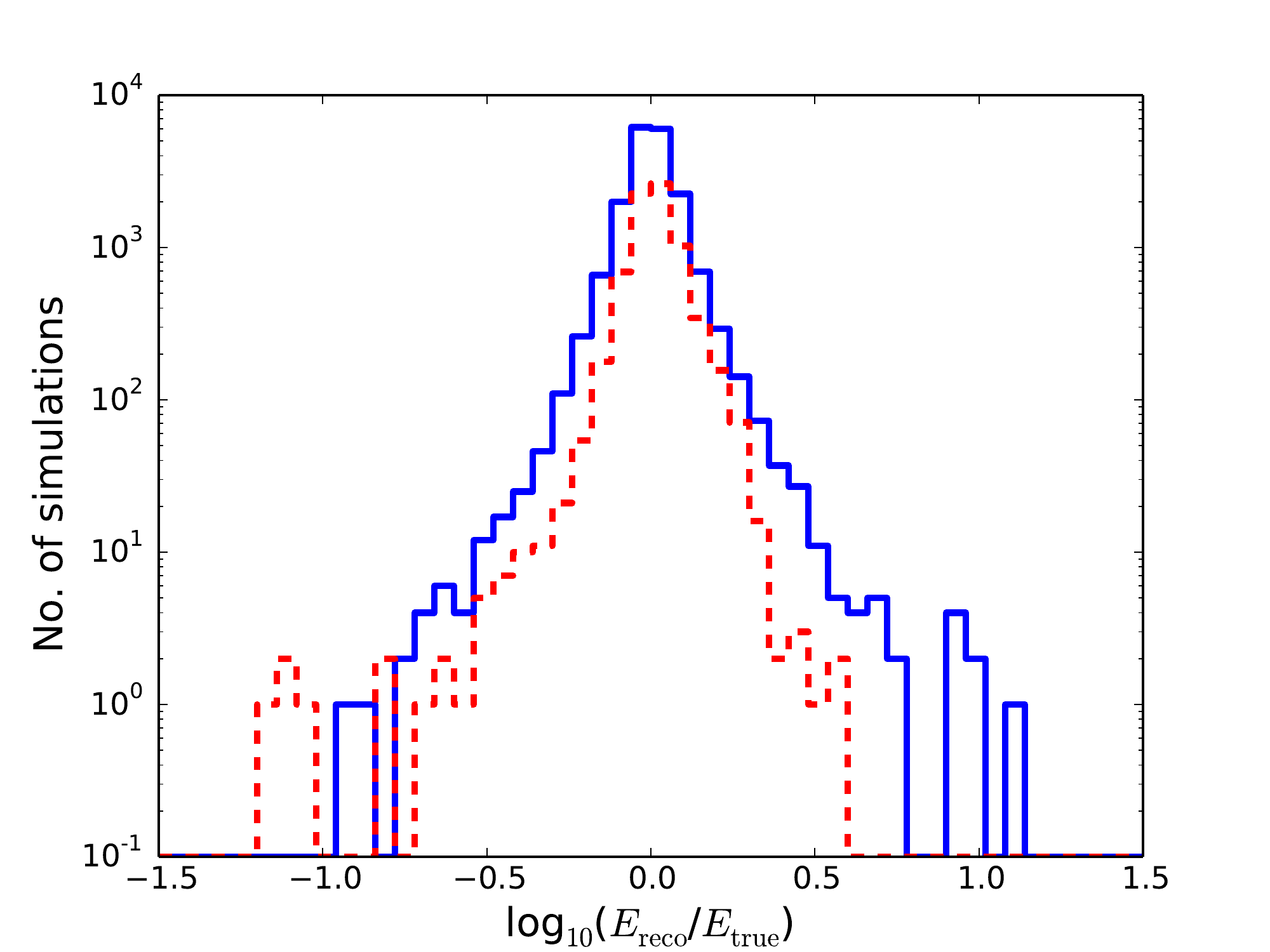}
\caption{{\bf Extended Data Figure 7 $|$ Energy reconstruction} Distributions of the ratio between true and reconstructed energy for proton (blue, solid) and iron showers (red, dashed). The two types of showers have a systematic offset of the order of  $\sim 1\%$.}
\end{figure}

\newpage

\subsection{Methods}
\subsection{Event selection}
Cosmic ray detection at LOFAR continuously runs in the background during astronomical observations. When 16 out of the 20 scintillator stations of the LORA particle array detect a signal, a trigger is issued and the ring buffers of all active antennas within a $\sim ~1$ km radius are stored for offline analysis\cite{pipeline}. Which antennas are active depends on the settings of the astronomical observation. For this analysis we have selected showers that were measured with at least 4 antenna stations (corresponding to at least 192 antennas) in the low band (30 - 80 MHz after filtering)

The trigger and selection criteria introduce a composition bias. This bias is removed with a cut based on dedicated sets of simulations that are produced for each observed shower.  These sets contain 50 proton and 25 iron showers, that span the whole range of possible shower depths. A shower is only accepted when \emph{all} simulations in its set pass the triggering and selection criteria. This anti-bias cut removes many showers below $10^{17}$~eV, but only 4 above that energy. In this analysis, we restrict ourselves to the higher-energy showers and impose a cut on energy $E_{\mathrm{reco}} > 10^{17}$~eV.

The energy cut itself is another potential source of compositional bias, as the reconstructed energy might be dependent on the depth of the shower. However, in our reconstruction approach this effect is very small because energy and $X_{\mathrm{max}}$ are fitted simultaneously. Extended Data Figure 7 shows distributions of the ratio between true and reconstructed energy for proton and iron simulations. The systematic offset between the two particle types is of the order of $\sim 1$\%.

We used data from the Royal Netherlands Meteorological Institute to check for lightning storm conditions during our observations. When lightning strikes have been detected in the North of the Netherlands within an hour from a detection, the event is flagged and excluded from the analysis. The presence of electric fields in the clouds can severely alter the radio emission even in the absence of lightning discharges\cite{buitink07}. The polarisation angle of the radio pulse is very sensitive to the nature of the emission mechanism \cite{polarisation, polprl} and is used as an additional veto against strong field conditions.      

Finally, a quality cut is imposed on the sample in order to only include showers that have a core position and arrival direction that allows accurate reconstruction. We use the dedicated sets of simulations produced for each shower to derive uncertainties on core position, energy and $X_{\mathrm{max}}$. These three values are highly correlated, so a single cut on the core uncertainty of  $\sigma_{\mathrm{core}}<5$ m is sufficient.

The quality cut is based on the dedicated sets of simulations. These sets are produced for a specific combination of core position and arrival direction. Therefore, the quality cut is effectively a cut on position and direction, and does not introduce a composition bias. 

Furthermore, we stress that there is no cut on the quality of the reconstruction of the actual data. By applying the cuts described above we obtain a sample of 118 showers that are fitted to the simulation yielding reduced $\chi^2$-values in the range 0.9-2.9. Deviations from unity can be ascribed to uncertainties in antenna response, atmospheric properties like the index of refraction, or limitations of the simulation software.   

\subsection{Reconstruction}
The energy and $X_{\mathrm{max}}$ of the shower are reconstructed with the technique described in Buitink et al.\ (2014)\cite{Buitink2014}.

\subsection{Statistical uncertainty}
The statistical uncertainty on the power measurements of individual antennas include three contributions. First, there is contribution from the background noise which is a combination of system noise and the Galactic background. Secondly, there is a contribution from uncertainties in the antenna response model. There can be differences between the responses of antennas, either because of antenna properties (e.g. cross-talk between nearby antennas), or because of signal properties (e.g. polarisation). Since these fluctuations are different for each shower core position and arrival direction, they are essentially random and included as a 10\% statistical uncertainty on the power. A third contribution is due to the error introduced by interpolating the simulated pulse power. Strictly speaking this is not a measurement uncertainty, but it must be taken into account when fitting the data to simulation. The interpolation error is of the order of 2.5\% of the maximum power\cite{Buitink2014}. The three contributions are added in quadrature and produce the one sigma error bars shown in Extended Data Figures 1-5.     

The statistical uncertainty on $X_{\mathrm{max}}$ is given by the quadratic sum of the uncertainties due to reconstruction technique and the atmospheric correction. The former is found by applying our analysis to simulated events with added Gaussian noise, where the noise level is determined from the data.
  
In the CORSIKA simulations the standard US atmosphere model was used. The reconstructed shower depth is corrected for variations in the atmosphere using data from the Global Data Assimilation System (GDAS) of the NOAA National Climatic Data Center. We follow a procedure developed by the Pierre Auger collaboration\cite{gdasauger}. This typically leads to adjustments of the order of 5--20 g/cm$^2$. Remaining uncertainty after correction if of the order of 1 g/cm$^2$

The index of refraction of air is a function of temperature, air pressure, and relative humidity. Using local weather information the final data sample was split in two equal size groups corresponding to conditions with relatively high or low index of refraction. The mean reconstructed $X_{\mathrm{max}}$ of these groups deviate from the mean by $\pm 5$ g/cm$^2$, and we adopt this value as an additional statistical uncertainty. Because the refractivity used in simulation corresponds to dry air there is also an associated systematic error (see below).

The total statistical uncertainty on $X_{\mathrm{max}}$ is found by adding above factors in quadrature. A distribution of the uncertainty for the showers in our final sample is shown in Extended Data Figure~6.
 
The energy resolution is 32\% and is found by comparing energy scaling factors of the radio power and particle density fit (see Figure 1). 

\subsection{Systematic effects}
The data has been subjected to several tests to find the systematic uncertainty on the reconstructed values for $X_{\mathrm{max}}$:
\begin{itemize}
\item{\bf Zenith angle dependence} The final data sample is split into two groups of equal size by selecting showers with a zenith angle below or above 32 degrees. For both groups the mean reconstructed $X_{\mathrm{max}}$ is calculated, yielding deviations from the mean value of $\pm 8$ g/cm$^2$. This spread is larger than expected from random fluctuations alone and is included as a systematic uncertainty. The dependence on zenith angle may be related with atmospheric uncertainties (see below).  
\item{\bf Index of refraction of air} As explained above, the index of refraction changes because of differences in atmospheric conditions. Fluctuations on $X_{\mathrm{max}}$ due to changing humidity are of the order of 5 g/cm$^2$ with respect to the mean. However, the index of refraction that was used in the radio simulations corresponds to dry air, and is a lower bound to the actual value. Therefore, the real value of $X_{\mathrm{max}}$ can be higher than the reconstructed value but not lower, and we adopt an asymmetrical systematic uncertainty of $+10$ g/cm$^2$. 
\item{\bf Hadronic interaction model} Since the reconstruction technique is based on full Monte Carlo simulations, it is sensitive to the choice of hadronic interaction model that is used. It has been shown with a comparison between QGSJETII.04, SYBILL 2.1, and EPOS-LHC, that the uncertainty due to model dependence is $\sim 5$ g/cm$^2$. Note that the uncertainty on the \emph{composition} due to different models (in other words: on how to interpret the measured $X_{\mathrm{max}}$ values) is of course larger. 
\item{\bf Radiation code} For this analysis we have used the radiation code CoREAS in which the contributions of all individual charges to radiation field are added together. The advantage of this microscopic approach is that it is completely model-independent and based on first principles. ZHAireS\cite{zhaires} is another microscopic code and gives very similar results\cite{convergence}. To calculate the emission CoREAS uses the end-point formalism\cite{endpoint}, while ZHAireS is based on the ZHS algorithm\cite{ZHS}. Both formalisms are derived directly from Maxwell's equations and have been shown to be equivalent\cite{SLAC}. 
The other difference between CoREAS and ZHAires is that they take the particle distribution from different air shower propagation codes (CORSIKA and AIRES respectively) that internally use different hadronic interaction models. Since the radiation formalisms themselves are equivalent, small differences between CoREAS and ZHAireS are most likely due to differences in the hadronic interaction models used to simulate the particle interactions. The choice of radiation code does therefore not introduce an additional systematic uncertainty on top of the uncertainty to hadronic interaction models that is already included. A comparison study with LOFAR data did also not show any evidence for a systematic offset between the codes and will be published in an upcoming paper.   
\end{itemize}
The remaining small dependence of $X_{\mathrm{max}}$ on zenith angle is possibly related to the index of refraction. Showers with different inclination angles have their shower maximum at different altitudes, and therefore different local air pressure and index of refraction. Therefore, increasing the index of refraction used in simulations will result in a zenith-dependent change in reconstructed $X_{\mathrm{max}}$. This possibly removes the observed dependence of the composition on zenith angle. Correctly taking into account a complete atmospheric model for the profile of the refractivity of air is subject of further study. Here, we treat the effect conservatively by adding the first two contributions to the uncertainty linearly. The other two contribution are independent and are added in quadrature, yielding a total systematic uncertainty of $+14/-10$ g/cm$^2$.

The systematic uncertainty on the energy reconstruction with the LORA particle detector array is 27\%, which includes effects due to detector calibration, hadronic interaction models, and the assumed slope of the primary cosmic-ray spectrum in the CORSIKA simulations\cite{LORA, LORAanalysis}

\subsection{Statistical analysis}
For each observed shower, we calculate:
\begin{equation}
a = \frac{\langle X_\mathrm{proton}\rangle - X_\mathrm{shower}}{\langle X_\mathrm{proton}\rangle - \langle X_\mathrm{iron}\rangle}
\label{eq}
\end{equation}   
where $X_\mathrm{shower}$ is the reconstructed $X_{\mathrm{max}}$, and $ \langle X_\mathrm{proton}\rangle$ and $ \langle X_\mathrm{iron}\rangle$ are mean values predicted by QGSJETII.04\cite{qgsjet}. Thus $a$ is an energy-independent parameter that is mass sensitive. A pure proton composition would give a wide distribution of $a$ centered around zero, while a pure iron composition gives a narrower distribution around unity.

From the measurements we construct a cumulative distribution function (CDF) in the following Monte Carlo approach. A realisation of the data is made by taking the measured values for the energy and $X_{\mathrm{max}}$, adding random fluctuations based on the statistical uncertainty of these parameters, and calculating the $a$ parameters and the corresponding CDF. By constructing a large number of realisations with different random fluctuation, we can calculate the mean CDF and the region that contains 99\% of all realisations. These are indicated in Figure 3 as the solid blue line and the shaded region respectively.

We fit theoretical CDFs based on composition with two or four mass components to the data. The test statistic in the fit is the maximum deviation between the data and the model CDFs. The p-value is given by the probability of observing this deviation, or a larger one, assuming the fitted composition model. 

We first use a two-component model of proton and iron nuclei, where the mixing ratio is the only free parameter. The best fit is found for a proton fraction of 62\%, but it describes the data poorly with a p-value of $1.1\times 10^{-6}$.

A better fit is achieved with a four-component model (p+He+N+Fe), yielding a p-value of 0.17. While the best fit is found for a Helium fraction of 80\%, the fit quality deteriorates only slowly when replacing helium by protons. This is demonstrated in Figure~\ref{fig:contour} where the p-value is plotted for four-component fits where the fractions of helium and proton are fixed, and the ratio between N and Fe is left as the only free parameter. The solid line in this Figure contains the parameter space where $p>0.01$. We construct a 99\% confidence level interval on the total fraction of light elements (p+He) by finding the two extreme values of this fraction that still lie within the $p>0.01$ region.

The total fraction of light elements (p+He) is in the range [0.38,0.98] at 99\% confidence level, with a best fit value of 0.8. The heaviest composition that is allowed within systematic uncertainties (see above) still has a best fit p+He fraction of 0.6, and a 99\% confidence level interval of [0.18, 0.82].

\end{document}